\documentclass[conference]{IEEEtran}
\IEEEoverridecommandlockouts
\usepackage{cite}
\usepackage{amsmath,amssymb,amsfonts}
\usepackage{graphicx}
\usepackage{textcomp}
\usepackage{xcolor}
\usepackage{listings}
\usepackage{algorithm}
\usepackage{algpseudocode}
\usepackage{hyperref}
\hypersetup{hidelinks}
\usepackage{flushend}
\usepackage{svg}
\usepackage[shortlabels]{enumitem}
\usepackage{multirow}
\usepackage{balance}
\usepackage{booktabs}

\def\BibTeX{{\rm B\kern-.05em{\sc i\kern-.025em b}\kern-.08em
    T\kern-.1667em\lower.7ex\hbox{E}\kern-.125emX}}

\definecolor{codegreen}{rgb}{0,0.6,0}
\definecolor{codegray}{rgb}{0.5,0.5,0.5}
\definecolor{codepurple}{rgb}{0.58,0,0.82}
\definecolor{backcolour}{rgb}{0.95,0.95,0.92}

\lstdefinestyle{mystyle}{
    commentstyle=\color{codegreen},
    keywordstyle=\color{magenta},
    numberstyle=\tiny\color{codegray},
    stringstyle=\color{codepurple},
    basicstyle=\ttfamily\footnotesize,
    breakatwhitespace=false,         
    breaklines=true,                 
    captionpos=b,                    
    keepspaces=true,                 
    numbers=left,                    
    numbersep=5pt,                  
    showspaces=false,                
    showstringspaces=false,
    showtabs=false,                  
    tabsize=2
}

\ifCLASSOPTIONcompsoc
    \usepackage[caption=false,font=normalsize,labelfont=sf,textfont=sf]{subfig}
\else
    \usepackage[caption=false,font=footnotesize]{subfig}
\fi

\newif\iffinal

\finaltrue

\iffinal
    \newcommand\ian[1]{}
    \newcommand\kyle[1]{}
    \newcommand\dejan[1]{}
    \newcommand\alok[1]{}
    \newcommand\valerie[1]{}
    \newcommand\andre[1]{}
    \newcommand\maxime[1]{}
\else
    \newcommand\ian[1]{{\color{red}[Ian: #1]}}
    \newcommand\kyle[1]{{\color{green}[Kyle: #1]}}
    \newcommand\dejan[1]{{\color{cyan}[Dejan: #1]}}
    \newcommand\alok[1]{{\color{orange}{Alok: #1}}}
    \newcommand\valerie[1]{{\color{blue}{Valerie: #1}}}
    \newcommand\andre[1]{{\color{purple}{Andre: #1}}}
    \newcommand\maxime[1]{{\color{brown}{Maxime: #1}}}
\fi

\newcommand\sysname{\texttt{GreenFaaS}}
\newcommand\anoncluster{Institutional Cluster}

\begin{document}

\title{GreenFaaS: Maximizing Energy Efficiency of HPC Workloads with FaaS}

\author{
\IEEEauthorblockN{Alok Kamatar, Valerie Hayot-Sasson \\ Yadu Babuji, Andre Bauer}
\IEEEauthorblockA{Department of Computer Science\\
University of Chicago\\
Chicago, Illinois, USA \\
\{alokvk2,vhayot,yadunand,andrebauer\}\\@uchicago.edu}
\and
\IEEEauthorblockN{Gourav Rattihalli, Ninad Hogade\\Dejan Milojicic}
\IEEEauthorblockA{Hewlett Packard Labs\\
Santa Clara, California, USA \\
\{gourav.rattihalli, ninad.hogade, \\dejan.milojicic\}@hpe.com}
\and
\IEEEauthorblockN{Kyle Chard \\ Ian Foster}
\IEEEauthorblockA{Department of Computer Science\\
University of Chicago\\
Chicago, Illinois, USA \\
\{chard,foster\}@uchicago.edu}
}

\maketitle

\pagestyle{plain}

\begin{abstract}
    Application energy efficiency can be improved by executing each application component on the compute element that consumes the least energy while also satisfying time constraints.
    In principle, the function as a service (FaaS) paradigm should simplify such optimizations by abstracting away compute location, but existing FaaS systems do not provide for user transparency over application energy consumption or task placement.
    Here we present \sysname{}, a novel open source framework that bridges this gap between energy-efficient applications and FaaS platforms. 
    \sysname{} can be deployed by end users or providers across systems to monitor energy use, provide task-specific feedback, and schedule tasks in an energy-aware manner.
    We demonstrate that intelligent placement of tasks can both reduce energy consumption \textit{and} improve performance. For a synthetic workload, \sysname{} reduces the energy-delay product by 45\% compared to alternatives.
    Furthermore, running a molecular design application through \sysname{} can reduce energy consumption by 21\% and runtime by 63\% by better matching tasks with machines.
\end{abstract}


\section{Introduction}
As we can no longer expect rapid, significant improvements in the energy efficiency of hardware, there is a critical need for software solutions to mitigate application energy consumption~\cite{temporal-shifting}. 
One approach to reducing energy consumption is to schedule parts of workloads to more energy efficient devices~\cite{galantino2023assessing, wimpy-nodes}. 
The flexible scheduling of fine-grained tasks supported by the Function-as-a-Service (FaaS) model~\cite{baldini2017serverless,fox2017status} can, in principle, be helpful in this regard~\cite{hpe-cloud,enex,first}, but in practice the conventional FaaS model also abstracts physical hardware, making it impossible for users to monitor the energy used by applications, and leaving them at the mercy of FaaS providers to mitigate energy use. 

Meanwhile, existing tools for improving application energy efficiency deal only with local or single-machine deployments~\cite{func-energy-andre, geopm}, and thus are constrained in the improvements that they can achieve by the properties of a single machine. 
Furthermore, many of these tools do not monitor the energy consumed by tasks at runtime,
relying on static, offline energy use models instead~\cite{mhra, enex, hpe-cloud}.
Yet in heterogeneous multi-system environments, both energy consumption and performance can vary significantly with task-machine assignment. 
We demonstrate the potential for significant energy savings by better matching functions to machines, but find that realizing these opportunities requires an online monitoring framework and an automatic placement algorithm that accounts for the energy costs of both data transfer and task execution.

To address these needs, we present \sysname{}, a tool to monitor and schedule FaaS functions across machines.
In short, \sysname{}: (i) collects energy information from running FaaS tasks, (ii) accounts for the energy consumption of data transfers, and (iii) aggregates task and transfer energy consumption information and uses that information in deciding where to place tasks. \sysname{} also provides a web-based interface to increase user awareness of energy consumption and thus incentivize change~\cite{green-feedback}. 
\sysname{} can be used by providers, but also by end-users to more efficiently leverage existing systems without elevated privileges.
Our work represents a significant step towards empowering users to manage and reduce the energy footprint of their applications.
It also exposes the need for additional node, cluster, and network-level information to give users a more accurate picture of their energy use. 
\sysname{} is available on GitHub: \url{https://github.com/AK2000/caws}.

The contributions of our work are:
\begin{itemize}
    \item An analysis of energy use of FaaS functions deployed on a commodity desktop and  HPC machines, showing the need for online energy monitoring and the potential of multi-site scheduling to increase energy efficiency.
    \item The development of GreenFaaS, an open source FaaS scheduling framework that allows for online monitoring and energy-aware placement of FaaS tasks.
    \item A novel energy-aware scheduling algorithm to account for the distinct properties of FaaS tasks.
    \item A case study of a scientific application using GreenFaaS that shows a 63\% speedup and 21\% reduction in energy consumption.
\end{itemize}

The remainder of this paper is as follows. \autoref{sec:motivation} motivates GreenFaaS by profiling tasks across systems; \autoref{sec:design} presents our system design, prediction methodology, scheduling algorithm, and web interface; \autoref{sec:evaluation} evaluates \sysname{} and provides case study using a molecular design application; \autoref{sec:related_work} discusses related work; \autoref{sec:limitations} describes limitations and potential extensions of our work; and finally \autoref{sec:conclusions} concludes.

\section{Motivation}
\label{sec:motivation}
Traditionally, cloud/HPC users directly specified and configured an application for a specific site. This left little flexibility to relocate the application, or individual functions within that application, to different sites. FaaS lifts the level of abstraction away from specific systems, by providing a uniform interface for users to submit tasks without configuring them for a specific machine. Amazon GreenGrass extended the FaaS interface to allow functions to be run on IoT and edge devices. Globus Compute introduced a ``bring your own compute'' model, allowing users to run FaaS functions on their own infrastructure (see \autoref{sec:globus-compute}). Using these federated FaaS platforms enables users to easily choose to run certain tasks on any of the machines to which they have access. We hypothesize that this ability can be used to reduce the energy consumption of running an application without sacrificing performance by \textit{better} matching tasks to machines. To understand if this is possible and the requirements of such a system, we conduct experiments to investigate three important questions, namely:

\begin{enumerate}[start=1,label={\bfseries Q\arabic*:}]
\item How does the machine affect the performance and energy consumption of a task?
\item Do we need to explicitly collect energy consumption information for each task?
\item How should a user or provider decide where to run a task?
\end{enumerate}

\begin{table*}[ht]
    \centering
    \caption{Machines used in experiments. TDP=Thermal Design Power, in Watts. Idle power is for all sockets on the node.}
    \label{tab:systems}
    \begin{tabular}{rrrrrrrr}
        \toprule
         Machine & Year Deployed & CPU Model & \# of Cores & CPU TDP (W) & Idle Power (W) & Avg.\ Queue (s) \\
         \midrule
         Desktop & 2022 & Intel Core i7-10700 & 16 & 65 & 6.51 & 0 s \\
         ALCF Theta & 2017 & Intel KNL 7320  &  64 & 215 & 110 & 32 s \\
         \anoncluster{} (IC) & 2021 & 2 $\times$ Intel Xeon 6248R & 48 & 205 & 136 & 24 s\\
         TAMU FASTER & 2023 & 2 $\times$ Intel Xeon 8352Y & 64 & 205 & 205 & 22 s\\
         \bottomrule
    \end{tabular}
\end{table*}

\subsection{Testbed}
We run experiments on three HPC machines (Theta, \anoncluster{}, FASTER) and a personal workstation.
Theta is an older supercomputer at the Argonne Leadership Computing Facility (ALCF); IC is the Midway-3 Institutional Cluster at the University of Chicago; and FASTER is a Texas A\&M University (TAMU) resource available through the NSF's ACCESS allocations~\cite{access}. 
These four systems differ significantly in their generation, size, and architecture, as shown in  \autoref{tab:systems}.
In the table, 
\# of Cores is per node,
CPU Thermal Design Power (TDP) 
is the maximum heat, in Watts, that a single CPU is designed to dissipate;
Idle Power is the total power consumption (of all CPUs on a node) measured when a node is allocated and only running the monitoring code;
and Avg.\ Queue is the observed time between a Globus Compute endpoint requesting a single-node job (e.g., from the local batch scheduler) and that job starting, averaged across multiple runs.
We see that the three HPC systems have similarly powerful CPUs, while Desktop is about a third as powerful. This difference is also reflected in the measured idle power use. Desktop uses 6~W when not running a task, whereas each CPU on the HPC systems uses 100$+$W.

We deploy four Globus Compute endpoints, one per system, each configured to request a single node at a time and to deploy one worker per core on a node. 
We measure node energy consumption while tasks are running by using the Cray Power Monitor~\cite{cray-hss} on Theta and the RAPL interface on the other three systems.
We attribute energy consumption collected for each node to individual tasks as described in Section \ref{sec:energy-consumption}. 

We use functions from the Serverless Benchmark Suite (SeBS)~\cite{sebs} shown in \autoref{tab:functions}. Each such function is a single-core CPU function. 
Since container support varies across HPC systems, we preinstalled any code dependency on the system, and tasks were run without containerization. To collect the performance of each task on a machine, we run 1, 2, 4, etc., tasks, up to the number of cores, and record both the 
execution time and energy consumed.

\renewcommand{\arraystretch}{1.8}
\begin{table}[ht]
    \centering
    \caption{Serverless functions derived from the Serverless Benchmark Suite (except for matrix multiplication).}
    \label{tab:functions}
    \begin{tabular}{rrr}
         \toprule
         Function &  Description & Features \\
         \midrule
         Graph BFS & Python IGraph BFS & Graph Size \\
         Graph MST & Python IGraph MST  & Graph Size \\
         Graph Pagerank & Python IGraph Pagerank  & Graph Size \\
         Compression & Folder compression using tar & Folder Size \\
         DNA & Visualize with Squiggle & File Size \\
         Thumbnail & \parbox[r]{3.5cm}{\raggedleft Image size reduction using PIL} & File Size \\
         \parbox[r]{1.25cm}{\raggedleft Video Processing} &  \parbox[r]{3.5cm}{\raggedleft Conversion or water-marking using \texttt{ffmpeg}} & \parbox[r]{1.5cm}{\raggedleft File Size, Operation} \\
         \parbox[r]{1.6cm}{\raggedleft Matrix multiplication} & \parbox[r]{3.5cm}{\raggedleft Numpy matrix multiplication, double precision} & Data Size\\
         \bottomrule
    \end{tabular}
\end{table}

\subsection{Experiments and Discussion}

We conduct experiments to answer the three questions listed at the start of this section.

\begin{figure*}
    \centering
    \hfill
    \subfloat[Runtime]{
        \centering
        \includegraphics[width=.3\textwidth]{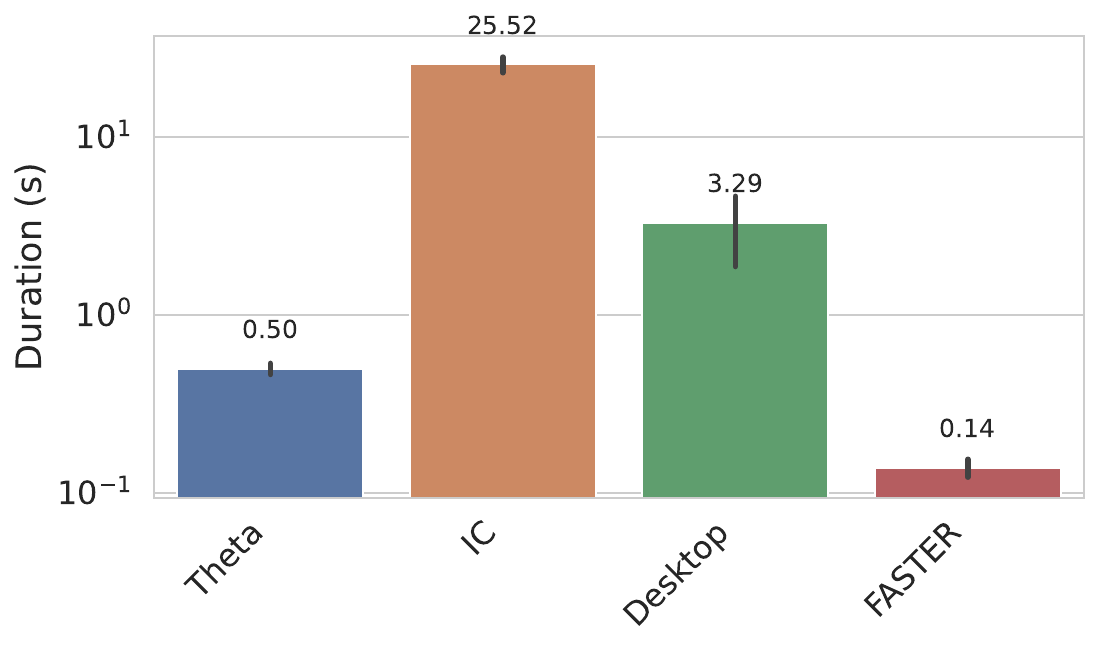}
    }
    \hfill
    \subfloat[Total energy]{
        \centering
        \includegraphics[width=.3\textwidth]{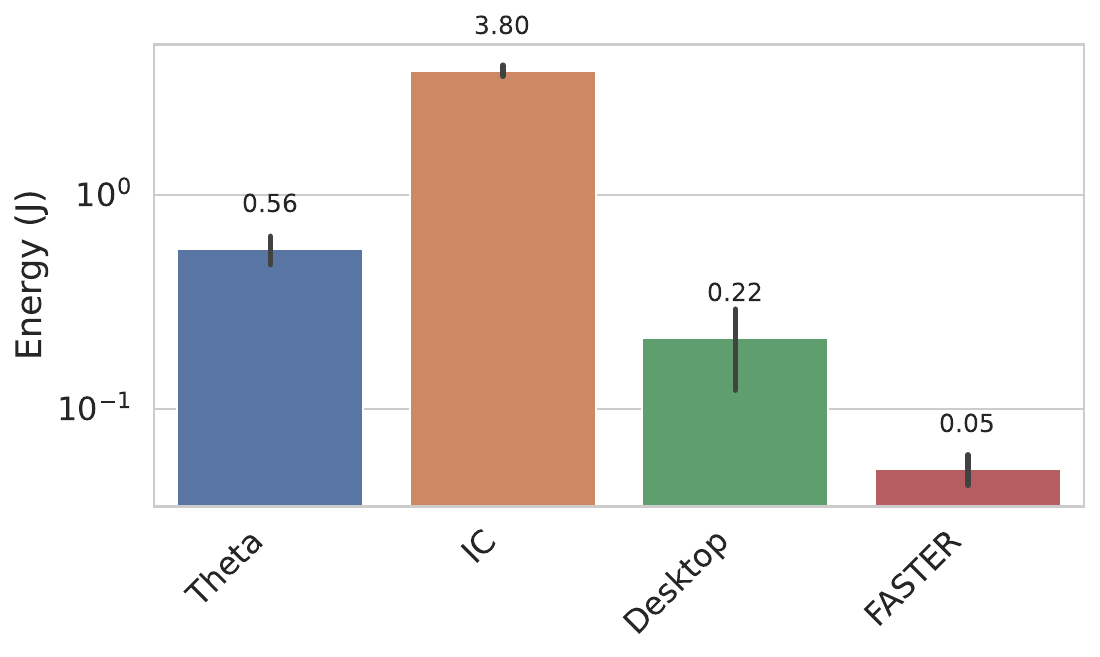}
    }
    \hfill
    \subfloat[Average power]{
        \centering
        \includegraphics[width=.3\textwidth]{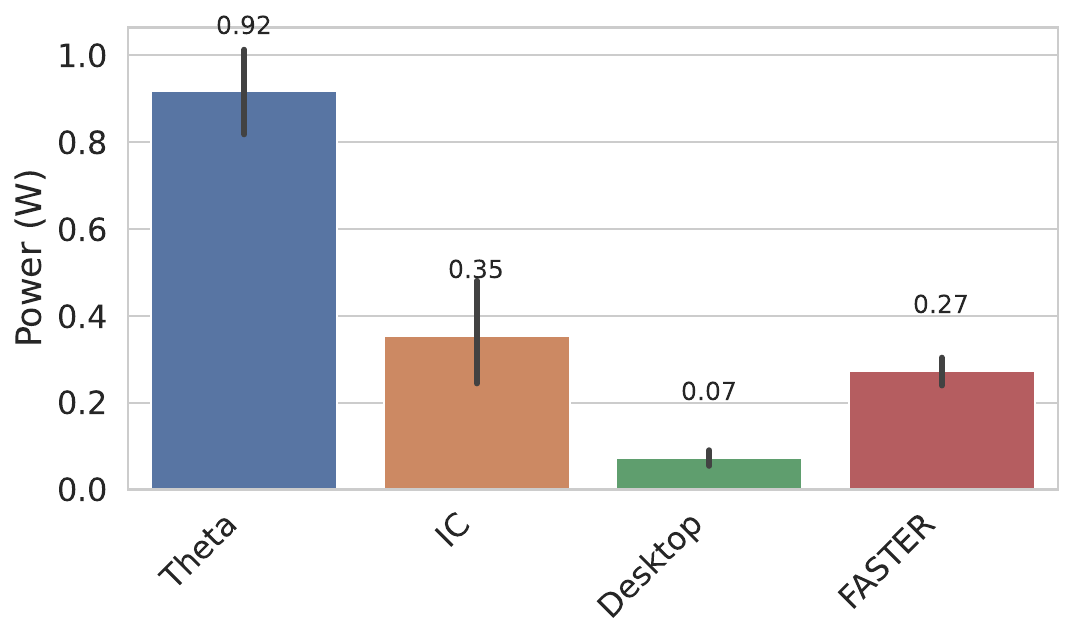}
    }
    \hfill
    \caption{Per-machine runtime, energy, and average power for the Graph Pagerank benchmark.}
    \label{fig:pagerank_comparison}
\end{figure*}

\emph{Q1: How does the system affect the performance/energy consumption of a task?} 
The potential savings of the choice of endpoint are not well understood. 
For instance, when comparing a lower-specification CPU to a higher-specification one (in terms of TDP, frequency, etc.), we expect a corresponding increase in processing speed or number of cores that could lead to similar energy efficiencies. 
For simplicity, we use a single function, the graph\_pagerank benchmark, to investigate this balance.
\autoref{fig:pagerank_comparison} shows runtime, energy, and power for this program on each of the four systems.
The fastest machine (FASTER) runs graph\_pagerank about 200$\times$ faster than does the slowest (\anoncluster{}), and consumes 
75$\times$ less energy, for a saving of 
3.7J if we consider only the incremental energy used by a task.
Including idle power draw complicates the picture. For Desktop, idle power is drawn whether or not we are running tasks, whereas on HPC machines, the batch scheduler could allocate an unused node to another job.  
Thus, if we consider idle power draw, for a single task Desktop is actually more efficient than FASTER, albeit with lower performance, introducing a trade-off between runtime and energy. 
FASTER only becomes more efficient if there are enough tasks so that the idle power draw can be amortized.
On the other hand, as long as Desktop has capacity, it always makes sense to migrate a graph\_pagerank task from \anoncluster{} to it, since it since it will always use less energy and run faster. 
\textbf{Key Takeaway: Optimal task placement can provide significant energy savings, potentially without sacrificing performance.}

\begin{figure*}
    \centering
    \subfloat[Runtime]{
        \centering
        \includegraphics[width=.3\textwidth,trim=2mm 4mm 2mm 2mm,clip]{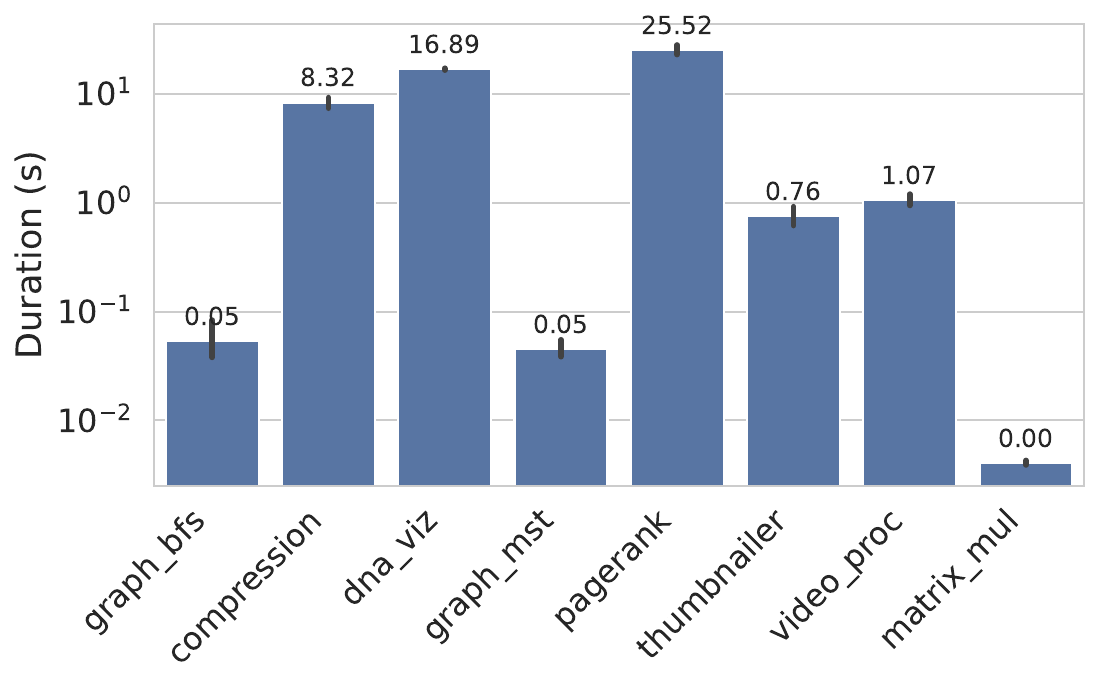}
    }
    \subfloat[Total energy]{
        \centering
        \includegraphics[width=.3\textwidth,trim=2mm 4mm 2mm 2mm,clip]{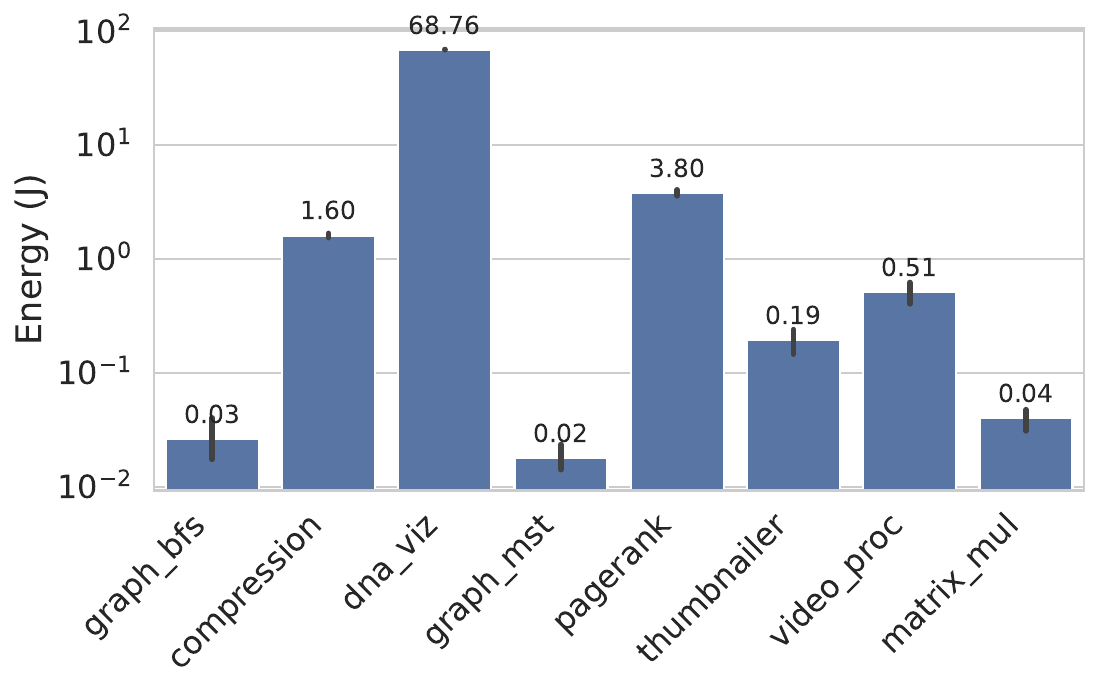}
    }
    \subfloat[Average power] {
        \centering
        \includegraphics[width=.3\textwidth,trim=2mm 4mm 2mm 2mm,clip]{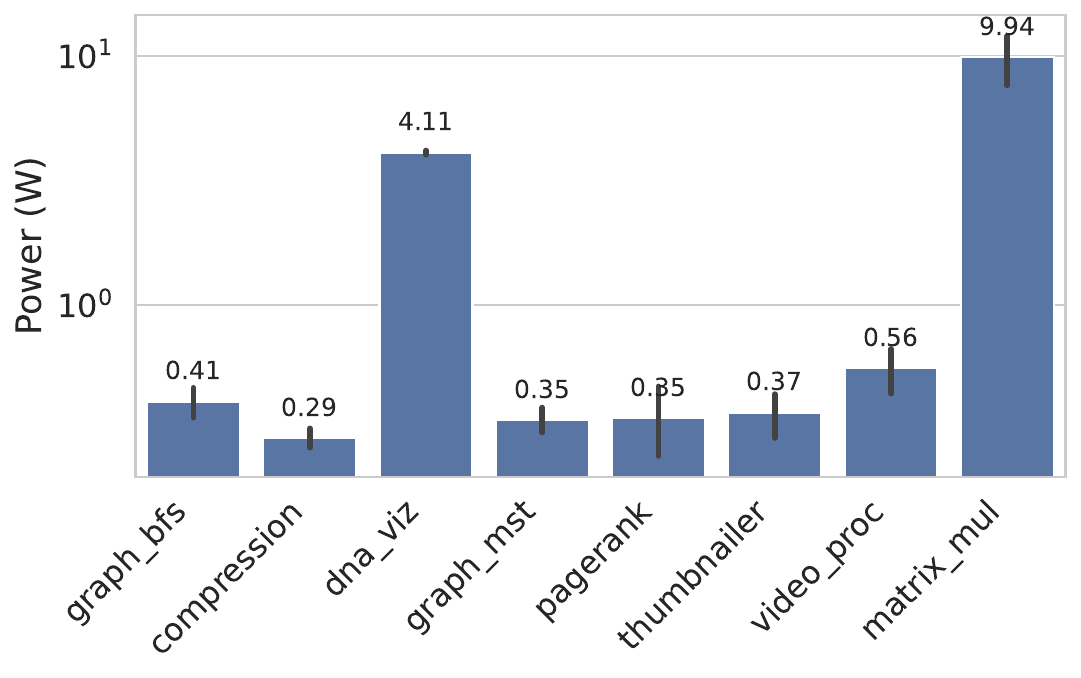}
    }
    \caption{Per-function runtime, energy, and average power for eight serverless benchmark suite functions on \anoncluster{}.}
    \label{fig:sebs_midway_profile}
\end{figure*}

\emph{Q2: Do we need to  collect energy consumption information explicitly for each task?}  The critical question here is whether we can estimate and reduce energy use without energy monitoring infrastructure being built into a FaaS system. Previous work has assumed that the power consumption of a system can be modelled effectively by the cores occupied~\cite{hpe-cloud, mhra, enex}, 
assuming that all tasks would behave the same way.
To investigate the accuracy of this assumption, we show the runtime, energy consumed, and average power for each benchmark running on \anoncluster{} in \autoref{fig:sebs_midway_profile}. 

In the figure, we see that there are still significant variations in the average power of a system that obscure the relationship between runtime and energy. For instance, on \anoncluster{}, even though dna\_visualization runs 10 seconds faster than graph\_pagerank, it consumes 18$\times$ more energy.  This difference in energy use would be missed without explicitly collecting energy information. There is also a discrepancy between systems on which tasks are energy-efficient.  For instance, although matrix\_mul has a lower average power than compression on FASTER (not shown), it uses 34$\times$ more power than compression on \anoncluster{}. 

It is difficult for us to determine the reasons why CPU power use may vary. The power consumed by a CPU can be roughly modelled by using its frequency~\cite{enex}, and operating systems/runtimes may use dynamic voltage-frequency scaling (DVFS) to optimize for energy consumption~\cite{heath2005energy}. Still, the degree to which frequency scaling affects application performance depends greatly on task-specific properties~\cite{first, murana-power-consumption}. We also examined whether IO-intensity (as measured by cache miss frequency) could explain the variation in power consumption across tasks on the same processor, or across processors. 
However, even if there is some underlying relationship between IO intensity and energy-use, we did not observe it in our measurements.
A task with a high frequency of cache misses on one system did not necessarily have a high frequency on another system, and even on a single system, IO-intensity was not well correlated with power use. 

These complexities make it difficult for users to profile and understand the power consumption of their software \textit{a priori}. Tools that assist in offline profiling are useful for developing green software~\cite{func-energy-andre} but are insufficient for deploying energy-efficient serverless functions, since power use can vary so greatly over machines. We conclude that energy monitoring should be integrated into FaaS platforms in order to collect information from the function execution environment and to build online task execution profiles. \textbf{Key Takeway: Online energy profiling of functions is necessary in order to determine energy-efficient task placement.}

\begin{figure}
    \centering
    \subfloat[]{
        \includegraphics[width=\linewidth,trim={0 0 0 0cm},clip]{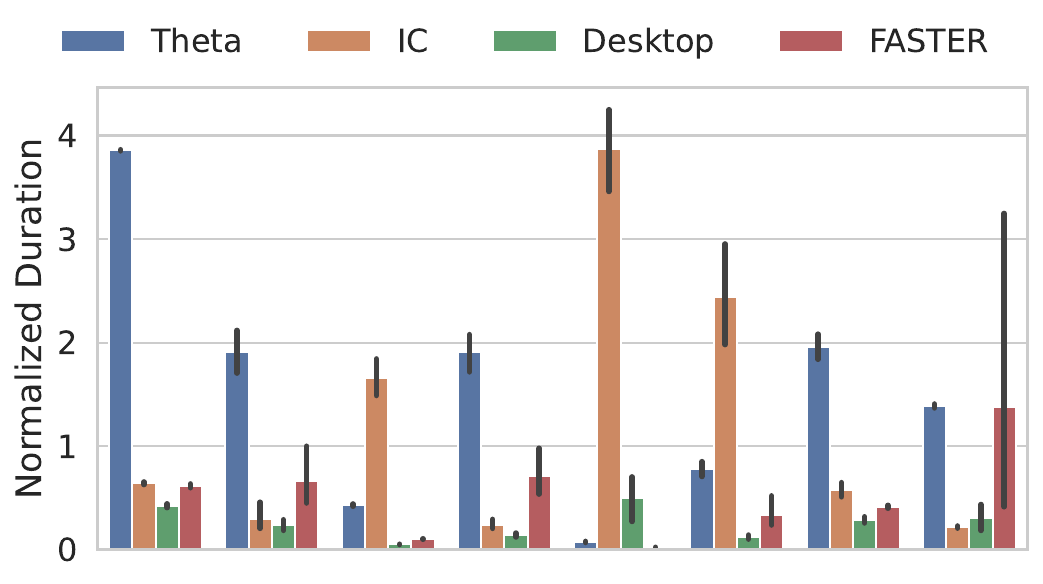}
    }

    \vspace{-8mm}
    
    \subfloat[]{
        \includegraphics[width=\linewidth,trim={0 4mm 0mm 1.4cm},clip]{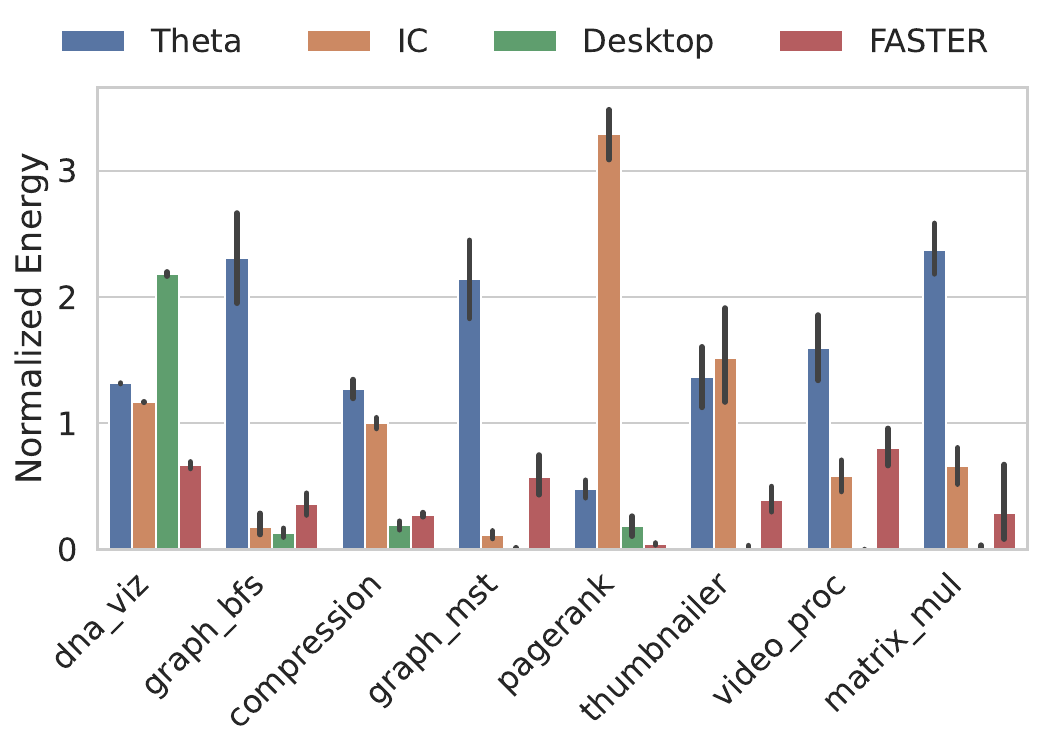}
    }
    \caption{Runtime and energy comparison of benchmark tasks across the four systems.
    Each values is normalized to the average for the corresponding task across all systems.}
    \label{fig:sebs_comparison}
\end{figure}

\emph{Q3: How should a user or provider decide where to run a task?} Given Q1 and Q2, we can now ask how a user or FaaS provider should decide where to run a task. Benchmarks such as Green500~\cite{green500} and SPECPower Benchmarks~\cite{spec} distill the performance of a machine to a single number, giving the impression that there is an optimal machine that should always be used if possible. If this were the case, tasks could easily be scheduled manually by users or via a simple heuristic. 

To examine this problem space, in \autoref{fig:sebs_comparison}, we compare the runtime and power draw for each task on each machine, normalized to the average across machines for that task. We find that no one machine is the fastest or the most efficient for all of the tasks, and that every machine uses less time or energy than average for at least one of the tasks.
Furthermore, other constraints must be taken into account such as queue time, idle power draw, or capacity. For instance, while relatively efficient and fast, Desktop is constrained to 16 cores compared to 48--64 for the other machines, limiting the number of tasks that can be sent to it before the wait times start to dominate application performance.  For relatively simple applications with a homogeneous bag of tasks, users may be able to select the optimal machine manually based off prior monitoring. However, when an application involves many tasks or a complex dependency graph, manually analyzing the trade-offs between machines quickly becomes infeasible.
\textbf{Key Takeaway: Users/providers need an automatic scheduling algorithm to ease the burden of efficient task placement.}

\section{System Design}
\label{sec:design}

\begin{figure}[t]
    \centering
    \includegraphics[width=\linewidth,trim= 16mm 9mm 8mm 6mm,clip]{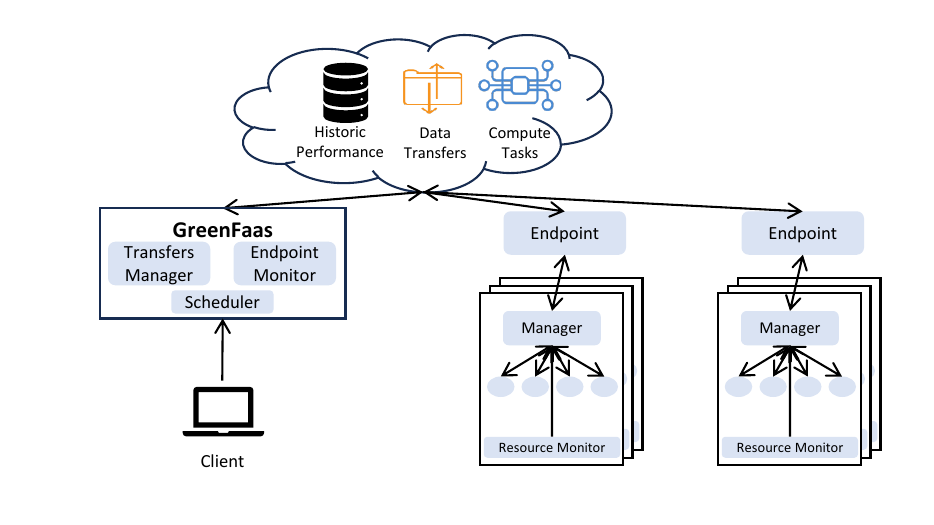}
    \caption{\sysname{} high-level architecture,  showing integration with Globus Compute and Transfer.}
    \label{fig:sys_overview}
\end{figure}

We next describe the \sysname{} system, focusing on how it collects energy information, manages data transfers and estimates their energy consumption, and uses this information to schedule workloads. 

\subsection{Overview}
Our goal is to provide the energy information that can allow users to a) make informed decisions when deciding where to execute functions and b) automate the selection of endpoints based on predicted energy consumption. 
We require that our approach: 1)~work on 
existing systems with differing hardware, software, and policies; 2)~run in user space without elevated permissions; 3)~avoid extensive offline profiling of hardware or software; 4)~require no user modifications to the FaaS service; and 5)~provide human-readable feedback on the energy consumption of their tasks.

\subsection{Background on Globus Compute}
\label{sec:globus-compute}
We implement \sysname{} on the Globus Compute FaaS platform (formerly FuncX)~\cite{funcx}. This federated FaaS system allows system administrators and users to turn any machine into a function serving platform by deploying
an \textit{endpoint}.
Underlying each endpoint is a configurable provider/launcher adapted from Parsl~\cite{parsl} that supports dynamic provisioning and management of resources from various HPC schedulers. Users can then submit function invocations (tasks) from any (authenticated) client to the cloud-hosted Globus Compute service, which offers persistent storage of tasks, routes tasks to user-specified endpoints, and handles queuing of waiting tasks and completed results. 
In this work, we modify the open-source SDK (used for submitting tasks) and endpoint software. All changes remain compatible with the hosted Globus Compute service.  

\subsection{Collecting Monitoring Information}
To begin, we need a mechanism to collect resource and energy utilization from endpoints. 
Since FaaS tasks are often short~\cite{azure-faas-trace,Bauer2023Globus}, we require the ability to collect metrics at fine-grained intervals, on the order of a second. Further, we must deal with system policies; for example, most compute nodes on HPC systems do not have outbound access to connect to a central database to save collected metrics.

Our monitoring architecture captures user-space metrics within the Globus Compute endpoint and therefore can be installed trivially by the endpoint owner.
When a compute node is allocated, we start an additional resource monitoring process that periodically polls the system for newly created processes. For each process, the resource monitor collects a predefined set of hardware performance counters by using the perfmon library~\cite{perfmon}. 
Prior work has established the accuracy of using performance counters to estimate process-level power consumption while maintaining low overhead~\cite{smart-watts, func-energy-andre, proc-power}. Performance counter data are sent back to the endpoint, where they are forwarded to a cloud-hosted \sysname{} database.  To avoid creating more communication channels or relying on the shared file system, monitoring messages piggyback on the existing result communication mechanism from compute node to endpoint (typically deployed on a login node). When a task executes on a worker, a wrapper around the task sends information about the execution start and end time as well as the process id of the worker so that it can be integrated with the measured resources.

Different machines offer different methods of measuring the power used by the node depending on the installer/integrator, the CPU, and the available devices. To overcome these differences, we developed an energy monitor abstraction that can be configured per endpoint. The abstraction includes the ability to stack and compose arbitrary monitors to account for various devices on the system. We have implemented three different energy monitors. The RAPL energy monitor uses the Running Average Power Limiter sysfs interface on Intel and AMD CPUs to measure the total package energy~\cite{intel-manual}. Latter works have demonstrated its accuracy compared to in line power meters~\cite{rapl-in-action}. On systems installed/integrated by Cray, the Cray energy monitor leverages sysfs special files created by the Cray Hardware Supervisory System, which polls performance counters at 10 MHz and makes them available to user-space applications~\cite{cray-hss}. Finally, the Nvidia GPU energy monitor uses the Nvidia Management Library (NVML) to measure the GPU power consumption~\cite{nvml}. It can be composed with either of the other two monitors to measure both CPU and GPU energy.

\subsection{Estimating Energy Consumption}
\label{sec:energy-consumption}
To provide task level feedback, we must decompose the energy consumption measured from a node to individual processes and tasks. Following prior work~\cite{mhra}, we model the power consumption of a node at time $t$, $P_n(t)$ as:

$$ P_n(t) \approx \sum_R f_R(X_{R}),$$

\noindent
where each $R$ is a discrete resource (CPU package or GPU) and $X_R$ are performance counters related to that resource, and $f_R$ is a learned relationship. For CPUs, we collect the \texttt{LLC\_MISSES}, \texttt{INSTRUCTIONS\_RETIRED}, \texttt{CPU\_CYCLES}, and \texttt{REF\_CYCLES} hardware performance counters from the linux \texttt{perf} interface. Using the measured power from the energy monitor interface described above, we train a power model each device:
$$ P_{R} \approx f_R(X_R).$$

In line with previous work, we fit a linear model to the collected data~\cite{func-energy-andre, smart-watts}. 
A linear model allows us to easily decompose the power consumption into per-process measurements: 
$$f_R(X_R) = W_R \cdot X_R + B_R = (\sum_i W_R \cdot X_R^i) + B_R,$$
where $X_R^i$ are the performance counters for process $i$, $W_R$ is the learned weight matrix, and $B_R$ is the estimated idle power consumption calculated by fitting the model. We then say that the power consumption for process $i$ is $P_R^i = W_R \cdot X_R^i$. The security policies of HPC systems typically disallow kernel profiling and profiling of any process not owned by the user. We expect that the average system power use is accounted for by the constant $B$. However, since tasks can trigger system events that will not be accounted for in the power estimation, this results in an undercounting of performance events and estimation of task power. To account for this discrepancy, we adopt a correction factor to allocate the total measured power proportionally to the estimated power~\cite{smart-watts}:
$$\hat{P}_R^i = \frac{P_R}{W_R \cdot X_R} P_R^i.$$

This correction assumes that measured power not accounted for by the model is proportional to the estimated power. While this is an approximation, it allows us to account for system activities without elevated permissions. Note, that since a process is a software level abstraction, it is impossible to directly collect a ground truth measurement to evaluate the accuracy of this attribution. Others have evaluated similar models indirectly for attributes like consistency~\cite{decomposable-power, self-watts}.

Once we have determined the power consumption per process, we can attribute that power to individual tasks by computing the integral of the estimated power of the worker process from the task's recorded start time to its recorded end time. We use linear interpolation to account for high-frequency tasks, where the task sampling interval is a significant portion of task runtime. 

\subsection{Managing Data Transfers}
When federating multiple sites, data transfers need to be coordinated so that an endpoint on which a function is to be executed has access to required data.
We use Globus Transfer~\cite{chard14efficient}
to transfer files between sites as it supports third-party and high-performance transfers.
We assume that input files are passed as input arguments to the function and are annotated with the Globus Transfer endpoint at which they can be accessed. 
The annotation also serves to inform \sysname{} whether the file should be treated as task exclusive, or is to be shared between tasks on an endpoint. This latter feature allows files that are used by multiple tasks to be cached on an endpoint. The execution framework extracts these arguments and schedules required transfers before a task is executed. Applications can also return output paths annotated with the endpoint on which a task was executed, providing a mechanism for files to flow between tasks. To amortize overheads and avoid per-user Globus limits on concurrent transfers, data transfers for multiple tasks are batched before transfer.

To predict the transfer time of files between endpoints, we build a regression based on historical performance, where the features we consider are the number of files and the total size of the transfer. Because of the batching, the predicted transfer time cannot be calculated individually for each task, but only once all scheduling decisions in the batch have been made.

Estimating the energy use of file transfers is much trickier. There are the traditional complexities of predicting network performance without access to, or knowledge of, the hardware that is being used outside of the endpoints.  Even on-site, the energy use of a transfer is not transparent. On a typical HPC cluster, Globus endpoints are deployed on an exclusive Data Transfer Node (DTN). Once a file is received on the DTN, the shared file system data servers and metadata nodes take over transferring that file to the compute node. These components are shared among all users of the system and are access restricted so we cannot retrieve resource information from them. 

Given the barriers, we adopt a simplified model of energy used by data transfers. Offline, we measure the number of hops between endpoints using \texttt{tracert} (adding an additional hop for the shared file system and DTN, each, if applicable). We then model the energy consumption of a transfer from $n_1$ to $n_2$ as in prior work~\cite{ian-bits}:
$$E_{n_1 \rightarrow n_2} = \sum_h  s \times E^h_{inc},$$
where $h$ is the number of hops, $s$ is the transfer size in bytes, and $E_{inc}$ is the incremental power required to transmit a bit of data calculated by 
$$E_{inc} = \frac{P_{max}}{B},$$
where $P_{max}$ is the maximum power of the network device and $B$ is the bandwidth. For $P_{max}$ and $B$, we assume that each transfer engages core routers, edge routers, and switches, and choose specifications of typical network infrastructure matching those devices. 

\subsection{Scheduling}
\label{sec:sched}
Given a list of tasks $T = \{t_1...t_n\}$ and endpoints $E  = \{ e_1...e_m\}$ we are looking for a schedule $S: T \rightarrow E$ to achieve two goals: reduce energy consumption and improve performance (runtime). Since we expect the heterogeneity between different machines (in terms of runtime and energy consumption) to be larger than the differences within a machine, to simplify scheduling decisions, we assume that each endpoint implements its own placement algorithm to assign tasks to workers~\cite{hpe-cloud, first}. 

As there has been extensive work on energy-efficient task placement and scheduling, we build upon the state-of-the-art Multi-Heuristic Resource Allocation (MHRA) scheduling algorithm. MHRA was developed for energy efficient online task placement within heterogeneous data centers~\cite{mhra}. Since it has been shown to improve greatly the efficiency of cloud workflows---an analogous problem to the function placement problem---we choose here to adapt MHRA to the FaaS setting.
The resulting scheduling procedure is presented in \autoref{alg:scheduling}. MHRA defines an objective function that balances cost and runtime:
$$O = \alpha \frac{E_{tot}(S)}{SF_1} + (1-\alpha)\frac{C_{max}(S)}{SF_2}, $$
where $E_{tot}(S)$ is the total estimated energy consumption for a schedule $S$, $C_{max}(S)$ is the end time of the last task (makespan) of the schedule, $SF_1$ and $SF_2$ are normalizing constants, and $\alpha$ is a parameter to trade-off between energy efficiency and runtime. To get $SF_1$ and $SF_2$ we calculate the total runtime and energy of the (batch of) tasks being scheduled as if they were run on a single machine. This gives a pessimistic estimate for the value of runtime and energy. We treat $\alpha$ as a hyperparameter to the scheduler that gives the user control over the energy-runtime trade-off based on their requirements and explore the effects of $\alpha$ in our experiments. Then, we calculate the total energy $E_{tot}$  as:
$$E_{tot} = \sum_{n \in N} \int_{t_{n_{start}}}^{t_{n_{end}}} P_n(t) + \sum_{n_1 \in N} \sum_{n_2 \neq n_1} E_{n_1 \rightarrow n_2},$$
where $n,n_1, n_2 \in N$ represent the set of machines used. The first term  $\sum_{n \in N} P_n(t)$ is the total energy across all machines and $\sum_{n_1 \in N} \sum_{n_2 \in N} E_{n_1 \rightarrow n_2}$ is the total estimated cost of transfers between all pairs of machines. 

For $t_{n_{start}}$ and $t_{n_{end}}$ we use the start of the first task, and the estimated completion time of the last task, on node $n$, plus additional overhead to startup and release the node. That is, we consider only the energy consumed when the node is allocated to our workload. For endpoints without a batch scheduler (e.g., one running on a desktop), we consider the endpoint power for the entire span of the workflow.

MHRA proceeds by first ordering incoming tasks according to a heuristic (e.g., longest task first, highest average energy consumption first), line 7 of \autoref{alg:scheduling}. Then, starting with an empty schedule, the algorithm makes a greedy scheduling decision for each task. That is, for each task, the algorithm tries each machine $e$ (line 12), and calculates the value of the objective function for $S'$  as if the task were scheduled on that machine (line 14). It chooses the schedule $S'$ with the minimum objective across all possible machines (line 16). The algorithm repeats the process for different heuristic orderings (line 6), creating a different schedule for each heuristic (line 20). Then, it then returns the schedule with the lowest overall objective across all of the heuristics (line 22). Here, we consider Shortest/Longest Runtime First, and Highest/Lowest Energy Consumption First as different possible heuristics. 

Given that HPC nodes have a high idle power consumption compared to that of a single task, we find the greedy decision (line 12) almost never allocates tasks to a new node, even when the cost could be amortized over many tasks. To overcome this deficiency, we represent each task as a vector of energy and runtime predictions, $t^{m_1}_r$, $t^{m_1}_e$, which is the predicted runtime and energy use respectively of task $t$ on machine $m_1$. The predictions are an average of historical performance of that function on machine $m_1$. We then use agglomerative clustering to group similar tasks until the energy consumption of a cluster is greater than the energy consumed to start a node (line 5).
The high-level idea is to amortize the cost of allocating/starting up a new node across the cluster, while not changing the energy-runtime trade-offs between systems. For instance, if we are running many graph\_pagerank tasks that are more efficient on Desktop than on Theta (as shown in \autoref{fig:pagerank_comparison}), the tasks are put into a cluster that has this same property. The original greedy resource allocation algorithm is then applied by cluster. We refer to this modified scheduling algorithm as Cluster MHRA.

\begin{algorithm}[t]
\caption{Cluster MHRA}
\label{alg:scheduling}
\begin{algorithmic}[1]
\State $E = \textnormal{Array}(|T|, |M|)$ \hfill \Comment{Create task embedding matrix}
\For{$t \in T$}
    \State $e_t \gets \begin{bmatrix}t^{m_1}_r & t^{m_1}_e & ... & t^{m_{|M|}}_r & t^{m_{|M|}}_e \end{bmatrix}$  
\EndFor
\State $C \gets \textnormal{AgglomerativeCluster}(T, E)$ \Comment{Apply task clustering.}

\For{$k \in H_{heuristic}$} \hfill \Comment{Try each heuristic}
    \State Sort $C$ by $k$
    \State $S \gets \{\}$ \hfill \Comment{For each cluster}
    \For{$c \in C$} 
        \State $S \gets \{(c, E_0)\}$ 
        \State $f(S) \gets \alpha \frac{E_{flow}(S)}{SF_1} + (1-\alpha)\frac{C_{max}(S)}{SF_2}$
        \For{$m \in M$} \hfill \Comment{Try each endpoint}
            \State $S' \gets \{(c, m)\}$
            \State $f(S') \gets \alpha \frac{E_{flow}(S')}{SF_1} + (1-\alpha)\frac{C_{max}(S')}{SF_2}$
            \If $f(S') < f(S)$ 
                \State $S \gets S'$ \hfill \Comment{Make greedy decision}
            \EndIf
        \EndFor
    \EndFor
    \State $S_k \gets S$
\EndFor
\State \Return $\min_k(S_k)$ \hfill \Comment{Choose best heuristic}
\end{algorithmic}
\end{algorithm}

\subsection{Web Interface}
\begin{figure}
    \centering
    \includegraphics[width=\linewidth]{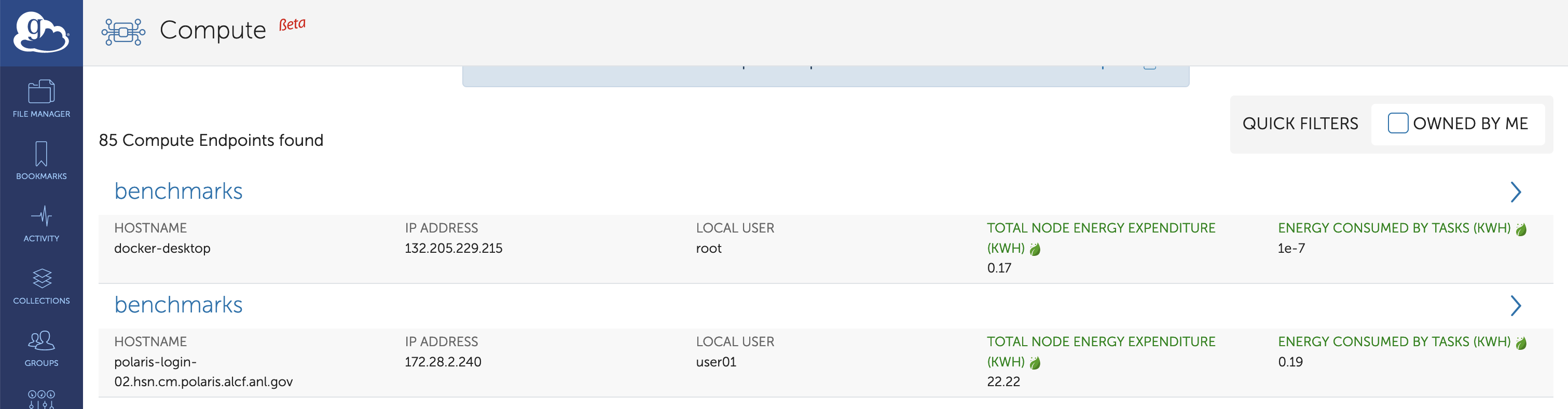}
    \caption{Globus web app  with the bookmarklet enabled.}
    \label{fig:bookmarklet}
\end{figure}

Without access to energy monitoring information, users remain unaware of the energy efficiency of their tasks and the endpoints on which those tasks run. To provide this information to users, we developed a ``bookmarklet'' to augment the Globus web application with endpoint energy usage information, as seen in \autoref{fig:bookmarklet}.

A bookmarklet is a web browser bookmark that can include Javascript code that is executed when loaded. Thus, users can add the bookmarklet by registering the GitHub-hosted Javascript code directly in their browser. The bookmarklet is executed when the page is loaded and the Javascript code dynamically modifies the rendered HTML. Our \sysname{} bookmarklet makes requests to the \sysname{} database to retrieve energy usage information. 

The information displayed by the bookmarklet includes the total node energy expended during task execution and the total energy usage of the user's tasks. Using this information as a guide, users can preselect the best endpoints for their tasks given their endpoint energy usage history.

\section{Evaluation}
\label{sec:evaluation}
We evaluate the performance of \sysname{} in terms of  overhead incurred the scheduling performance.
We use the same experimental setup as in \autoref{sec:motivation}. All data is initially placed on the desktop.

\subsection{Overhead}
\renewcommand{\arraystretch}{1.1}
\begin{table}[t]
    \centering
    \caption{Overhead of monitoring by machine. RTT (Round Trip Time) refers to the time between when a user submits a function and when results are received.}
    \label{tab:monitoring_overhead}
    \begin{tabular}{rrrrcrr}
         \toprule
         Function & Tasks &\multicolumn{2}{c}{No Monitoring} & \phantom{ab} & \multicolumn{2}{c}{Monitoring}\\
         & & \multicolumn{2}{c}{RTT (s)} && \multicolumn{2}{c}{RTT (s)}  \\
         \cmidrule{3-4} \cmidrule{6-7}
         & & Mean & Std. && Mean & Std.\\
         \midrule
         No-op & 1 & 1.59 & 0.23 && 1.62 & 0.25 \\
         No-op & 512 & 14.23 & 0.25 && 14.59 & 0.15\\
         Matmul & 64 & 2.01 & 0.007 && 2.01 & 0.005\\
         \bottomrule
    \end{tabular}
    
\end{table}

We begin by showing that \sysname{} incurs minimal overhead when monitoring endpoints and scheduling workloads. 

\subsubsection{Monitoring Overhead}
To evaluate the overhead of monitoring
we configured an endpoint on Theta with none of \sysname{}'s energy monitoring modifications.
First, we measure the latency overhead of the endpoint with monitoring compared to the endpoint without monitoring for a single no-op task. Next, we stress the result delivery system by submitting 512 no-op tasks that return a ``Hello World!" string. This test is to measure the effect of delivering monitoring data on the same channel as the results. Finally, we assess if there is extra overhead on the CPU by submitting matrix\_multiplication tasks to saturate the cores on the endpoint. For each experiment, we ran 30 trials and reported the mean and standard deviation. The results, shown in \autoref{tab:monitoring_overhead}, show that, in all of the experiments, the monitoring endpoint does not impose significant overhead on the execution time.

\subsubsection{Scheduler Overhead}
\begin{table}[t]
    \centering
    \caption{Scheduling overhead of task placement.}
    \label{tab:sched_overhead}
    \begin{tabular}{rrrcrr}
        \toprule
        Strategy & \multicolumn{2}{c}{256 Tasks} & \phantom{ab} & \multicolumn{2}{c}{2048 Tasks} \\
        \cmidrule{2-3} \cmidrule{5-6}
         {} &  Time (s) & Per Task (ms) && Time &  Per Task \\
         \midrule
         Round Robin & 3.8e-5 & $\approx 0$ && 2.8e-4 & $\approx 0$ \\
         MHRA & 0.389 & 1.51 && 13.92 & 6.8\\
         Cluster MHRA & 0.065 & 0.26 && 0.674 & 0.33\\
         \bottomrule
    \end{tabular}
\end{table}

To evaluate the scheduler overhead, we measure the scheduling time for each strategy based on the size of batches that are scheduled. The results are shown in \autoref{tab:sched_overhead}. The table shows the average time for each strategy to schedule a batch of 256 tasks and a batch of 2048 tasks, evenly distributed across the benchmark tasks. Although Globus Compute does not impose any limits on function execution, these values span the concurrency limits of cloud providers~\cite{sebs}.  As a baseline, we compare the Cluster MHRA algorithm that we described in \autoref{sec:sched} to a naive Round Robin approach and the original MHRA algorithm~\cite{mhra}. We see that the Cluster MHRA algorithm is approximately 6$\times$ faster than the original MHRA algorithm in the experiment with 256 tasks, and scales linearly across the region of interest. This improvement is explained by the number of decisions required by the scheduling algorithm - MHRA makes a separate decision for each task, while Cluster MHRA makes a decision for every cluster. In this case, each cluster contained between 12 and 40 tasks. The amortized scheduling overhead is 0.10~s per task. For context, the queue time in our testbed is approximately 30~s, and the invocation overhead of Globus Compute is reported to be 109~ms per task for a warm endpoint~\cite{funcx}. Thus, the overhead imposed by the scheduler is unlikely to impede performance improvements.

\begin{figure}[t]
    \centering
    \includegraphics[width=.9\linewidth]{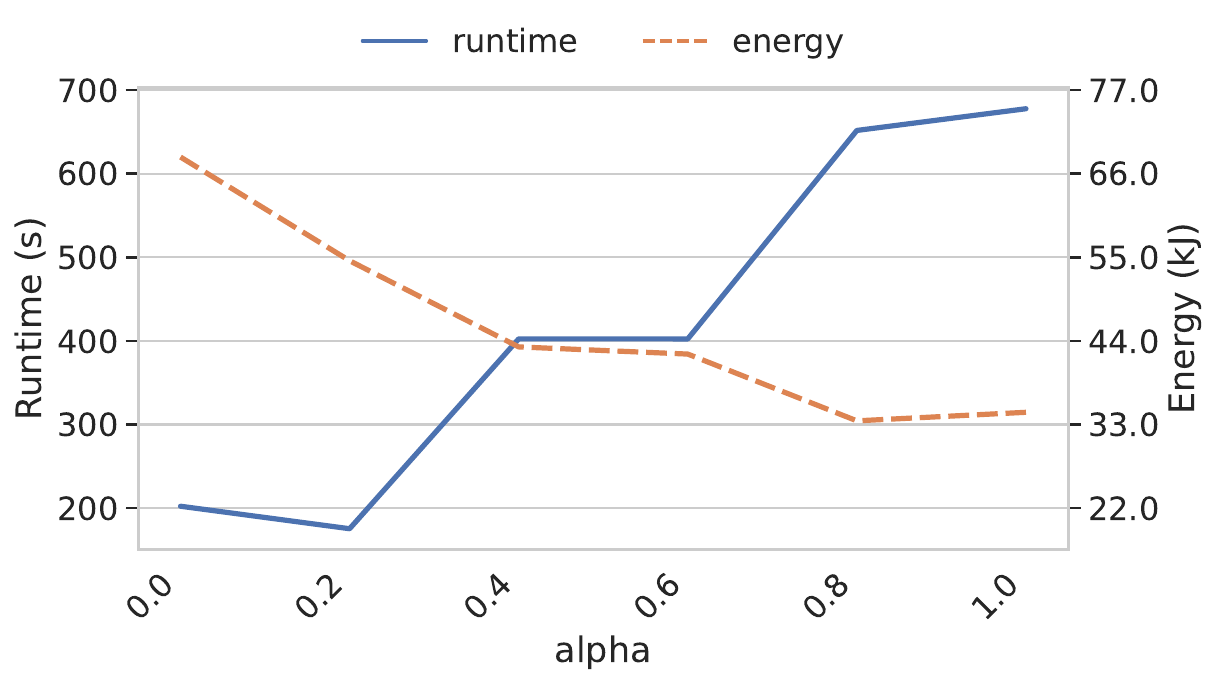}
    \vspace{-3mm}
    \caption{Sensitivity of scheduler to different values of $\alpha$, which determines the trade-off between energy and runtime. Higher $\alpha$ values prioritize energy over runtime.}
    \label{fig:sched_tradeoff}
\end{figure}
\begin{figure}[t]
    \includegraphics[width=.9\linewidth]{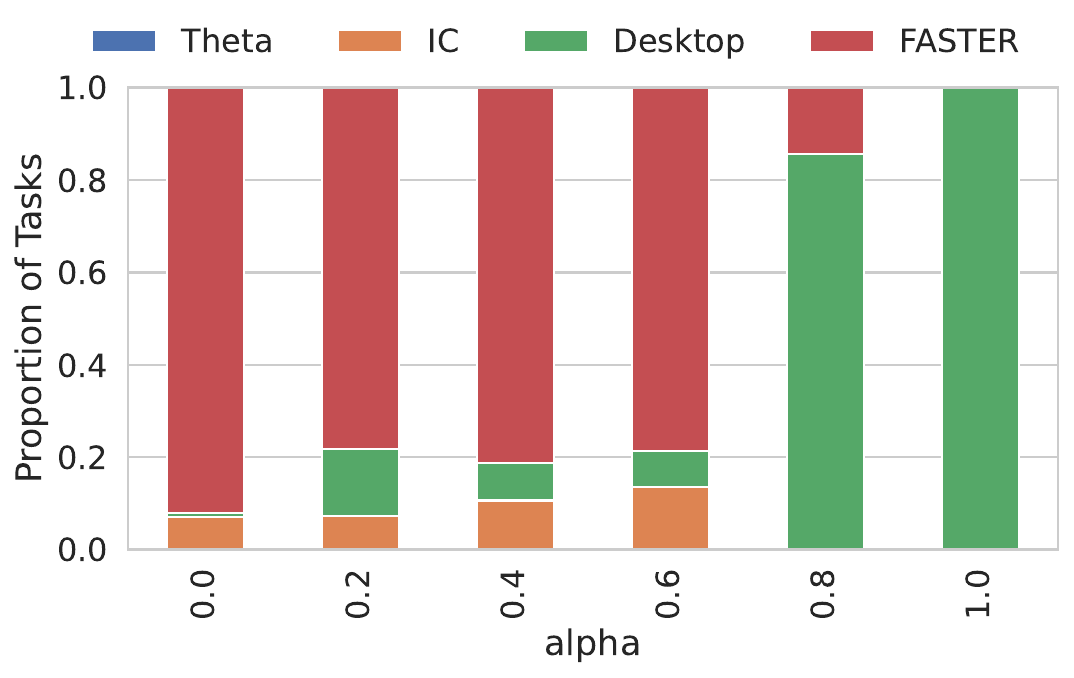}
    \vspace{-3mm}
    \caption{Task assignment distribution across values of $\alpha$.}
    \label{fig:sched_assignments}
\end{figure}

\subsection{Scheduler Evaluation}
\label{sec:sched_eval}
Lastly, we evaluate \sysname's task placement strategies using both benchmark FaaS tasks and a substantial scientific application. 

\subsubsection{FaaS Task Workload}
We define a sample FaaS workload comprising 256 invocations of each of the seven benchmarks, for a total of 1792 tasks. 
(Globus Compute places a 5MB limit on the size of invocations. The matrix multiplication benchmark caused Globus Compute to fail inconsistently because of this limit.)
We measure total runtime and energy consumed when this workload is submitted to \sysname{} for scheduling over our four experimental systems, while varying scheduling strategy.

First, we run the workload with our Cluster MHRA algorithm while varying $\alpha$ in the scheduling objective.
We see in \autoref{fig:sched_tradeoff} that as $\alpha$ goes from 0 to 1, the runtime of the placement strategy triples, while energy consumption is reduced by 50\%---illustrating how $\alpha$ can be used to balance energy and runtime. 
In \autoref{fig:sched_assignments}, we show the task assignments generated for different values of $\alpha$.
We see that for lower values of $\alpha$, more tasks are assigned to the FASTER machine and \anoncluster{} than for higher values of $\alpha$, when more tasks are assigned to the efficient Desktop machine---reducing energy consumption but increasing runtime. As end-users are able to use the scheduler directly, each user can set $\alpha$ based on their energy-consciousness and tolerance of delays.

We also conduct additional experiments, with results shown in \autoref{tab:sched_edp}, in which we submit all tasks to a single endpoint (rows 1--4); evenly distribute tasks among endpoints (row 5); and compare our Cluster MHRA (with two different $\alpha$ values; rows 7, 8) to an unmodified MHRA algorithm \cite{mhra} (row 6).
In addition to runtimes (which include the overhead of prediction and scheduling), total energy, total transfer energy, we report in each case the energy-delay product (EDP) and Weighted Energy Delay Squared Product (W-ED2P). 
EDP, calculated as energy $\times$ runtime, is a commonly used metric to examine trade-offs between circuit-level power-savings techniques \cite{edp}. W-ED2P is a modification of EDP tuned for HPC that more heavily weights the runtime \cite{Weighted-ED2P}. 

We see from \autoref{tab:sched_edp} that Desktop (the first row) is the most energy efficient endpoint, and running all tasks on it produces a schedule that uses the least energy. Cluster MHRA ($\alpha$=1.0; the second-to-last row) produces the same schedule and thus has similar energy use, except that it incurs additional overhead from scheduling. When we focus on the trade-off between runtime and energy, the benefits of our Cluster MHRA scheduling strategy are more apparent. Cluster MHRA with $\alpha$=0.2 reduces runtime by 16\% compared to the fastest alternative schedule (running all tasks on FASTER) and does so while reducing the energy consumption. The result is a 31\% improvement in EDP and a 42\% improvement in ED2P over the best alternative and 72\% improvement in EDP over the original MHRA. Furthermore, we see that without clustering, the original MHRA algorithm cannot balance runtime and energy use. Note that we show MHRA results for $\alpha = 0.5$ since varying $\alpha$ did not change the schedule produced by that algorithm.

\begin{table*}[t]
    \centering
    \caption{Comparison of task placement strategies. Transfer energy is an estimated value and not included in the Energy column. Energy-Delay Product (EDP) and Weighted Energy-Delay Squared Product (W-ED2P) are fused metrics to account for runtime and energy use. EDP and W-ED2P are normalized to the minimum for each column.}
    \label{tab:sched_edp}
    \begin{tabular}{rrrrrrr}
         \toprule
         Strategy & \centering Machine(s) & Runtime (s) & Energy (kJ) & Transfer Energy (kJ)  & EDP & W-ED2P  \\
         \midrule
         \multirow{4}{*}{Single node} & \centering Desktop & 640 & \textbf{33.5} & 0 & 2.24 & 11.7 \\
         & \centering Theta & 656 & 103 & 10.72 & 7.07 & 30.3 \\
         & \centering \anoncluster{} & 340 & 79.3 & 10.00 & 2.82 & 5.80 \\
         & \centering FASTER & 209 & 66.1 & 13.76 & 1.45 & 1.72 \\
         \midrule
         Round Robin & \centering All  & 272 & 69.6 & 8.72 & 1.98 & 3.19 \\
         \midrule
         MHRA & \centering All & 707 & 47.3 &  0 & 3.50 &  19.2 \\
         Cluster MHRA ($\alpha=1.0$)& \centering All  & 677 & 34.6 & 0 & 2.45 & 13.6\\
         Cluster MHRA ($\alpha=0.2$) & \centering All & \textbf{175} & 54.5 & 4.64 & \textbf{1.00} & \textbf{1.00}\\
         \bottomrule
    \end{tabular}
\end{table*}

\subsubsection{Molecular Design Application}

\begin{figure}
    \centering
    \def\svgwidth{\columnwidth}
\begingroup%
  \makeatletter%
  \providecommand\color[2][]{%
    \errmessage{(Inkscape) Color is used for the text in Inkscape, but the package 'color.sty' is not loaded}%
    \renewcommand\color[2][]{}%
  }%
  \providecommand\transparent[1]{%
    \errmessage{(Inkscape) Transparency is used (non-zero) for the text in Inkscape, but the package 'transparent.sty' is not loaded}%
    \renewcommand\transparent[1]{}%
  }%
  \providecommand\rotatebox[2]{#2}%
  \newcommand*\fsize{\dimexpr\f@size pt\relax}%
  \newcommand*\lineheight[1]{\fontsize{\fsize}{#1\fsize}\selectfont}%
  \ifx\svgwidth\undefined%
    \setlength{\unitlength}{187.5bp}%
    \ifx\svgscale\undefined%
      \relax%
    \else%
      \setlength{\unitlength}{\unitlength * \real{\svgscale}}%
    \fi%
  \else%
    \setlength{\unitlength}{\svgwidth}%
  \fi%
  \global\let\svgwidth\undefined%
  \global\let\svgscale\undefined%
  \makeatother%
  \begin{picture}(1,0.5)%
    \lineheight{1}%
    \setlength\tabcolsep{0pt}%
    \put(0,0){\includegraphics[width=\unitlength,page=1]{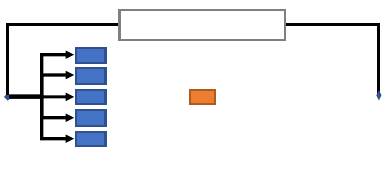}}%
    \put(0.3143688,0.424){\makebox(0,0)[lt]{\lineheight{1.25}\smash{\begin{tabular}[t]{l}Select Next Tasks\end{tabular}}}}%
    \put(0,0){\includegraphics[width=\unitlength,page=2]{workflow_svg-tex.pdf}}%
    \put(0.1184184,0.052){\makebox(0,0)[lt]{\lineheight{1.25}\smash{\begin{tabular}[t]{l}Simulate\end{tabular}}}}%
    \put(0,0){\includegraphics[width=\unitlength,page=3]{workflow_svg-tex.pdf}}%
    \put(0.3982868,0.052){\makebox(0,0)[lt]{\lineheight{1.25}\smash{\begin{tabular}[t]{l}(Re-)Train\end{tabular}}}}%
    \put(0,0){\includegraphics[width=\unitlength,page=4]{workflow_svg-tex.pdf}}%
    \put(0.749372,0.052){\makebox(0,0)[lt]{\lineheight{1.25}\smash{\begin{tabular}[t]{l}Infer\end{tabular}}}}%
    \put(0,0){\includegraphics[width=\unitlength,page=5]{workflow_svg-tex.pdf}}%
  \end{picture}%
\endgroup%

    \caption{Molecular design workflow~\cite{mol-des-github}}
    \label{fig:mol-des-workflow}
\end{figure}

Finally we evaluate the effectiveness of \sysname{} for measuring and optimizing energy use in a molecular design workflow. The application uses active learning to search for a molecule with the highest ionization energy. It consists of quantum chemistry simulation tasks, model training tasks, and inference tasks as shown in \autoref{fig:mol-des-workflow}. 
More details on this application are provided by Ward et al.~\cite{ward2021colmena}.

\begin{figure}
    \centering
    \includegraphics[width=0.9\linewidth]{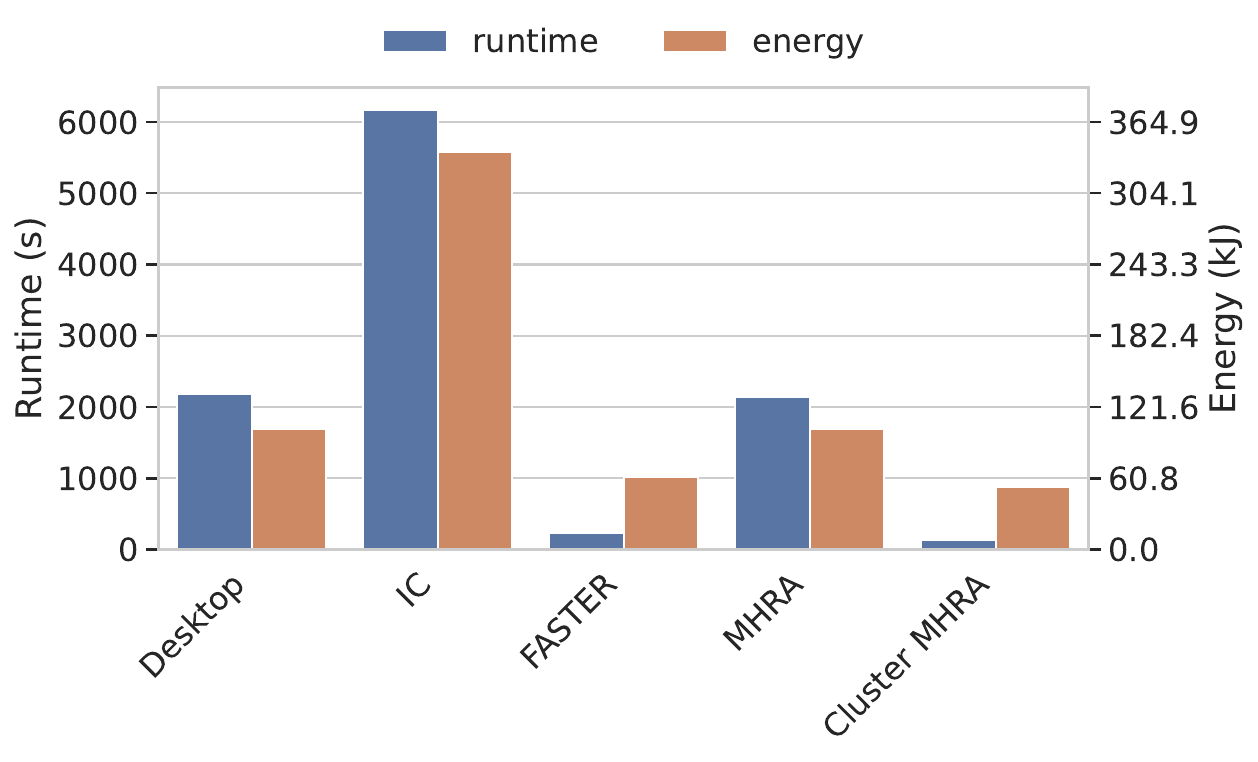}
    \vspace{-3mm}
    \caption{Runtime and energy consumed for molecular design application on three individual systems and when scheduled across all three systems with MHRA and Cluster MHRA.}
    \label{fig:mol-des-results}
\end{figure}

The application submits tasks to the FaaS scheduler only when they are ready to execute, so the scheduler does not know the full DAG ahead of time. The results are shown in \autoref{fig:mol-des-results}. 
(Theta was taken offline before these experiments were run.)
In this case, \sysname{} scheduling improves both the runtime and the energy efficiency compared to the best single site and MHRA. Specifically, running the molecular design application using \sysname{} and Cluster MHRA completes in 63\% less time, and consumes 21\% less energy than running the same workload on FASTER. Examination of the schedule produced by Cluster MHRA shows that this improvement comes by scheduling the highly-parallel simulation and inference stages on FASTER, while keeping the model training stage on Desktop, where it runs faster and uses less energy. 

\section{Related Work}
\label{sec:related_work}

\textbf{FaaS Scheduling.}
There has been little exploration of the energy consumption of FaaS workloads. EneX implements a scheduling algorithm for FaaS tasks based on integer linear programming \cite{enex}, but does not take into account task or processor heterogeneity. FIRST implements a meta-scheduling layer for FaaS platforms to minimize Optimal Operating Point divergence~\cite{first}. 
Such algorithms could be implemented within an endpoint to further reduce energy consumption. Galantino et al.\ assess the potential to save energy by distributing tasks in a workload across edge and cloud devices, similar to the Desktop vs.\ HPC site trade-off examined in this work, but do so only for a specific application~\cite{galantino2023assessing}. GreenCourier proposes distributing FaaS tasks based on the carbon intensity of the grid, but does not consider heterogeneity between machines and differences in energy consumption~\cite{chadha2023greencourier}. Other work considered FaaS scheduling to optimize for cost, runtime, or resource utilization \cite{matt-delta, fnsched, faas-deliver}.

\textbf{Energy and Power Aware Scheduling.}
Energy and power have long been a concern in both HPC and cloud environments.
Hsu and Feng pioneered energy reduction techniques in HPC systems by using dynamic voltage and frequency scaling (DVFS)~\cite{hsu2005power}. 
Numerous other works consider using DVFS to distribute power to meet QoS requirements~\cite{harada2006power}, optimize the placement of virtual machines~\cite{von2009power}, or reduce power off a workload's critical path~\cite{wang2010towards, workflow-deadline,app-aware-geopm}. Heath et al.\ consider distributing requests in a heterogeneous server cluster to optimize for throughput and energy use \cite{heath2005energy}.  Juarez et al.~\cite{mhra} first proposed the multi-heuristic resource optimization algorithm for task placement of scientific workflows in a cloud environment. However, they only use offline profiling to estimate energy.
Other authors investigate how to optimize power consumed for data transfers and network function virtualization~\cite{di2019cross,nine2023greennfv}.

\textbf{Energy/Power Monitoring.}
Other works address measuring and improving the power consumption of software. Feng et al.\ profile a scientific application running on a cluster and break down the energy use by component~\cite{feng2005power}, but rely on additional instrumentation to obtain measurements. 
Ramon et al.~\cite{decomposable-power} build per-component power models based on hardware performance counters, while Schmidt et al.~\cite{func-energy-andre} use performance counters to estimate power consumption and tracing to attribute that consumption to certain functions in software.
SmartWatts~\cite{smart-watts} and its successor SelfWatts~\cite{self-watts} are self-calibrating software power meters that provide per-process power estimation and are suitable for distributed resources. Our approach to monitor per-process energy is directly inspired by those works.  PowerJoular is a similar tool used to estimate software power consumption that also measures GPU energy use \cite{power-joular}. Noureddine et al.\  demonstrates that feedback on software power consumption increases user willingness to change their behavior~\cite{green-feedback}.

\section{Limitations and Future Work}
\label{sec:limitations}
\begin{figure}
    \centering
    \includegraphics[width=\linewidth,trim=0 0 0 25mm,clip]{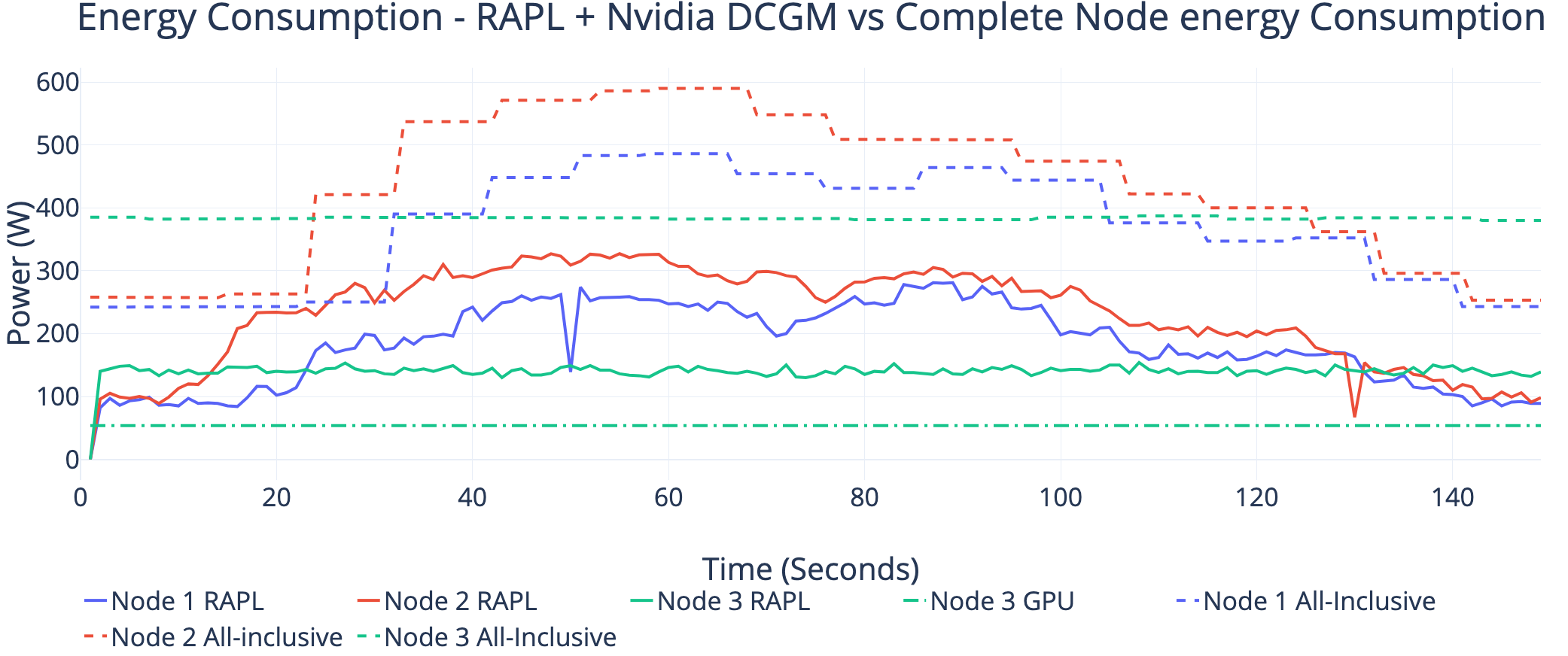}
    \caption{Comparison of energy information collected from RAPL, Nvidia DCGM, and Baseboard Management Controller (BMC) on three sample nodes. BMCs require additional support from resource providers and facilities but can provide additional energy information.}
    \label{fig:raplilo}
\end{figure}

We discuss limitations of our approach and opportunities for future work in this area.

\subsection{More Comprehensive Monitoring}
While RAPL provides fine-grained energy measurements for CPUs, it does not capture energy consumption by other node components, such as network interface cards, the cooling system or system-board/mother-board.  Even when requested, none of the three systems used in this work were able to provide node or whole system power usage information.  

One path forward would be to integrate data from Base Management Controllers (BMCs), small devices that are usually part of the system board and that permit active monitoring of a node's various metrics. As an example, \autoref{fig:raplilo} compares the data collected from BMCs on three sample nodes with that collected by using RAPL and Nvidia DCGM. 
All three nodes have 2 AMD EPYC 7443 CPUs, and node 3 has a Nvidia 80GB A100 GPU attached.
While the CPU-only RAPL measurements show similar trends to the whole-node BMC data, 
the figure highlights the impact of other energy costs on total resource energy consumption.
Monitoring using BMC information would provide a more accurate estimation of a workload but would require support from providers.

\subsection{Hierarchical Scheduling}
Monitoring additional components further increases the amount of data that must be transferred to, and processed by, the central scheduler. A potential solution is to combine multiple levels of energy-aware scheduling. A machine-level scheduler would be responsible for placement decisions within a machine and could manage fine-grain power allocation, and a higher-level framework like \sysname{} would decide which machine to run on. More detailed metrics would then need to be communicated only locally; the local agent would then communicate relevant information back to the global level. There is much prior work on which we could build that focuses on optimizing task placement or power distribution within a machine~\cite{app-aware-geopm, hpe-cloud, first}.

\subsection{Improved Incentives}
While increased energy efficiency can benefit resource providers by reducing costs, there is currently limited motivation for users to be more energy efficient.
Future work should look at methods for incentivizing users to be more energy conscious, such as pricing mechanisms based on energy use.

\section{Conclusions}
\label{sec:conclusions}

While in principle, the FaaS paradigm create compelling opportunities to reduce energy consumption, conventional FaaS platforms make it impossible for a user to monitor or reduce the energy use of their applications.
We have proposed \sysname{}, a system that is designed to improve the energy efficiency of FaaS workloads by monitoring energy consumption across sites, scheduling tasks to balance energy-runtime trade-offs, and providing information to users about their energy use.
Our results show that by using historical task information to better match tasks to machines, \sysname{} can speed up an application by 63\% while reducing energy consumption by 21\%. 
\sysname{} can be deployed on existing systems and allows users to track the energy impact of their application.
Such a system is critical to bridging the growing abstraction gap between applications and resources consumed, and to empowering users to write and run energy-efficient software in all environments.

\clearpage
\balance

\bibliographystyle{IEEEtran}
\bibliography{refs}

\end{document}